\documentclass[twocolumn,superscriptaddress]{revtex4-2}
\usepackage{xcolor}
\usepackage{graphicx}
\usepackage{dcolumn}
\usepackage{bm}
\usepackage{amsmath}
\usepackage{amsfonts}
\usepackage{multirow}
\usepackage[colorlinks=true, a4paper=true, pdfstartview=FitV,
linkcolor=blue, citecolor=blue, urlcolor=blue]{hyperref}

\newcommand{\bea}{\begin{eqnarray}}
\newcommand{\eea}{\end{eqnarray}}
\begin{document}
\definecolor{mygreen}{HTML}{006E28}
\newcommand{\ped}[1]{{\color{mygreen}\textbf{?PED:}  #1}}
\newcommand{\trom}[1]{{\color{red}\textbf{?TR:} \color{red} #1}}
\newcommand{\ya}[1]{\textbf{?YS:} {\color{blue} #1}}
\newcommand{\aw}[1]{{\color{magenta}\textbf{?AW:}  #1}}


\title{Spectral wall in collisions of excited Abelian Higgs vortices}

\author{A. Alonso Izquierdo}
\affiliation{
 Departamento de Matematica
Aplicada,  Universidad de Salamanca, SPAIN
}

\author{J. Mateos Guilarte}
\affiliation{
 Departamento de Matematica
Aplicada,  Universidad de Salamanca, SPAIN
}

\author{M. Rees}
\affiliation{School of Mathematics, Statistics and Actuarial Sciences, University of Kent, U.K}

\author{A. Wereszczy\'{n}ski}

\affiliation{
 Departamento de Matematica
Aplicada,  Universidad de Salamanca, SPAIN
}

\affiliation{
 Institute of Theoretical Physics,  Jagiellonian University, Lojasiewicza 11, 30-348 Krak\'{o}w, POLAND
}

\affiliation{International Institute for Sustainability with Knotted Chiral Meta Matter (WPI-SKCM2), Hiroshima University, Higashi-Hiroshima, Hiroshima 739-8526, JAPAN}

\begin{abstract}
We find a spectral wall in collisions of two vortices in the Abelian Higgs model at the critical coupling. It occurs if the out-of-phase mode of initially separated vortices is excited. 
\end{abstract}

\maketitle

\section{\label{sec:intro} Introduction}
In non-integrable field theories, topological solitons interact in a very complicated manner, whose comprehensive understanding is still a challenge. This issue is important not only from a conceptual or theoretical point of view but also is crucial for any realistic application of solitons. In general,  solitons interact in three qualitatively distinct ways. They can feel a static force from other solitons which leads to acceleration or deceleration of a soliton in multi-soliton environment. They can also interact via excitation of internal (vibrational) modes. Finally, radiation i.e., excitation of fluctuation modes in the continuum spectrum may affect their dynamics.

There is a particular class of models, called Bogomolny-Prasad-Sommerfield (BPS) models \cite{B, PS}, where dynamics is simplified due to non-existence of a static force between solitons \cite{Ma2}. Then, the simplest dynamics is captured by the geodesic flow through energetically equivalent static solutions of the corresponding Bogomolny equations. An example of such a process is a collision of vortices in the Abelian Higgs model \cite{R, Samols} or 
BPS monopoles in the Georgi-Glashow model \cite{Ma1} both at critical coupling.  

Surprisingly, it has recently been found that the geodesic dynamics is significantly affected if a bound mode, carried by the vortices, is excited. E.g. in a head-on 2-vortex scattering the famous single $90^\circ$ scattering is replaced by a chaotic (probably fractal) sequence of multi-bounces if the lowest mode (in-phase superposition of shape modes of each of 1-vortex) is excited \cite{Rees}. As explained in \cite{AMMW} excitation of this mode introduces an attractive force between the colliding single vortices. This, together with the resonant energy transfer mechanism between the kinetic and vibrational motion, triggers the fractal pattern of the multi-bounces, where depending on the number of collisions the vortices are scattered under $90^\circ$ or $180^\circ$ angle. Indeed, precisely as in the kink-antikink collisions in $\phi^4$ model \cite{sug, CSW, MORW}, energy initially stored in the kinetic motion can be temporary transferred into the vibrational mode and forces the vortices to collide again. 

Excitation of the upper mode (out-of-phase superposition of shape modes of each of the vortices) provides a repulsive vortex-vortex force and therefore cannot lead to multi-bounces. However, due to a level crossing with the third mode \cite{AMMW}, two bounces occur \cite{Rees}.

Here we will show that excitation of the upper mode in the 2-vortex collisions leads to the existence of another important surprise in solitonic dynamics, which is the {\it spectral wall} \cite{AORW}. The spectral wall (SW) phenomenon is an obstacle (barrier) in dynamics of a topological soliton due to the transition of a vibrational mode into the continuum spectrum. If the amplitude of this mode is sufficiently large the soliton is reflected by the SW while if it is small enough it can pass through the SW. For a particular value of the amplitude the soliton forms a long-living quasi-stationary state at a given spatial point, when the mode enters the continuum spectrum.  

Until now SWs were observed in (1+1) dimensional systems only, see e.g., \cite{SW-1, SW-2}. Here, for the first time, we find them in higher dimensions. 

\section{\label{sec:spectrum} Spectral structure and forces}
The Abelian Higgs model in (2+1) dimensions is defined by the following Lagrangian
\begin{equation}
L=\int dx_1dx_2 \Big[ -\frac{1}{4} F_{\mu\nu}F^{\mu \nu} +
\frac{1}{2} \overline{D_\mu \phi}\, D^\mu \phi -\frac{1}{8} (\overline{\phi}\, \phi-1)^2 \Big]
\, ,
\label{Lag}
\end{equation}
where $x_{1,2}=(x,y)$, $\phi$ is a complex scalar and $A_\mu$ is a $U(1)$ gauge field. The covariant derivative is $D_\mu\phi=(\partial_\mu - i A_\mu)\phi$ and 
$F_{\mu\nu}(x)=\partial_\mu A_\nu(x) - \partial_\nu A_\mu(x)$ is the field strength tensor. The coupling constants are chosen such that the masses of the matter and gauge field are equal. In this case, called critical coupling limit, static BPS vortices obey the Bogomolny equations \cite{B}
\begin{equation}
D_1\phi\pm i D_2\phi=0, \;\;\; F_{12} \pm \frac{1}{2} \left( \bar{\phi}\phi - 1\right)=0.
\end{equation}
This implies that they saturate a topological energy bound, $E=\pi |n|$, where $n \in \mathbb{Z}$ is the vorticity. This is a topological charge because it is the winding number associated to the map $\phi_\infty : S^1_\infty \to S^1$ from the unit sphere at infinity to the target space unit sphere. 

In the BPS scenario an $n$-vortex solution can be understood as a collection of $n$ 1-vortices located at arbitrary positions, $z_k = x_k+i y_k, i=1,\dots,n,$ which define a {\it moduli space} \cite{T}. They correspond to (multiple) zeros of the matter field and maximal magnetic field $F_{12}$. Independently of the point in the moduli space the energy does not change. There is no static vortex-vortex force.   

Thus, the simplest scattering of the vortices is just a transition through the energetically equivalent static solutions. It found a refined description as a geodesic flow on the corresponding moduli space.

In the case of 2-vortex a metric on such a space was found by Samols \cite{Samols}.  An important result is the $90^\circ$ scattering in a head-on collision of two single vortices. The vortices collide only once and the final state is $90^\circ$ rotated w.r.t. the initial state. For concreteness we assume that the vortices collide along $x$-axis. Thus their positions are $z=\pm d$ with $d\in \mathbb{R}_+$. ($d\in \mathbb{R}_-$ gives identical configurations). After the collision point, at $d^2=0$, they pass to $y$-axis. Thus $d$ becomes imaginary. For convenience we chose  $d \in i \mathbb{R}_-$. The distance between them is $|2d|$.

\begin{figure}
	\centering
	\includegraphics[height=5cm]{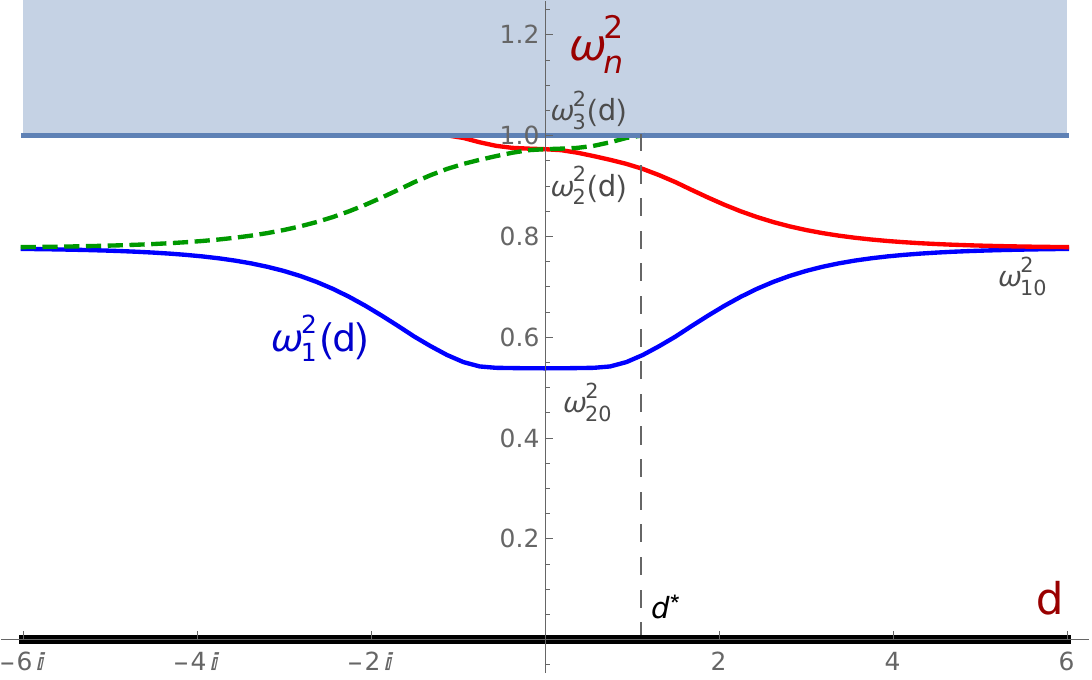} \hspace{1.0cm}
	\caption{The spectral structure for the 2-vortex solution as a function of the vortex position parameter $d \in \mathbb{R}_+ \cup i \mathbb{R}_-$ }  
	\label{fig:espectra}
\end{figure}

Although the energy of the $n$-vortex solution is degenerated on the moduli space, the spectral structure, that is, the structure of linear perturbations, changes. In particular, for the 2-vortex solution, even the number of the bound modes varies as we vary the distance between the constituent vortices \cite{AGMM}. For infinitely separated single vortices, $d \to \infty$, there are two degenerate bound modes. These are the in-phase and out-of-phase superposition of the radial shape mode of the 1-vortices \cite{H, AGG, AGG1}. Their frequencies are $\omega_{10}^2=0.777476$. As $d$ decreases (the vortices approach each other along the $x$-axis), the degeneracy is lifted and the in-phase (out-of-phase) superposition gives rise to the lower (upper) mode. The frequency of the lower (upper) mode monotonously decreases (increases) to $\omega_{20}^2=0.53859$ ($\omega_{21}^2=0.97303$) as the separation tends to 0, see Fig. \ref{fig:espectra}. 

Importantly, for $|d|<d_*$ the third shape mode exists. For $|d|\to d_*$ this mode approaches the mass threshold. As it becomes significantly wider close to this point, there is a rather big numerical uncertainty of the actual value of $d_*$. It was estimated in the range $d_* \in [1.2, 2.0]$ \cite{AGMM}. 

At $d=0$, where the vortices are on top of each other, the frequency of the lower mode takes a minimum value, while the frequencies of the two upper modes coincide. This degeneracy is in fact a mode crossing. Indeed, passing through $d=0$ the vortices change their location from $x$ to $y$-axis.  As shown in \cite{AMMW}, this means that as $d$ further decreases to negative imaginary values, the second mode of the 2-vortex transits into the third mode. Its frequency all the time grows.  

Decreasing of the frequency of the first mode results in appearance of an attractive inter-vortex force \cite{AMMW}. Now, the picture of a scattering is as follows. After the first $90^\circ$ collision the energy stored in the kinetic motion can be transfer into the internal energy of the mode vibrations. It can happen that the vortices have too little kinetic energy to overcome the attractive force. Thus they can collide again. The collisions may repeat several times until they receive sufficiently enough kinetic energy to escape to infinities. Such a resonant energy transfer leads to a fractal structure of multi-bounce windows. Note that, depending on the fact whether we have an odd or even number of collisions we get a $90^\circ$ or $180^\circ$ angle scattering. 

Analogously, the fact that the frequency of the second mode always increases, also after the coincidence point, amounts to the existence of a bit more nontrivial repulsive-attractive force. If this mode is initially excited the vortices repeal each other. Thus, for a sufficiently large mode amplitude a corresponding head-on collision may occur without passing through the on-top-of each-other configuration. The kinetic energy is simply too small to overcome the repulsion triggered by the upper mode. Due to that, there is no bounce and no $90^\circ$ scattering. For a bigger, fine tuned amplitude there can be one bounce (the vortices reach $d=0$), but still without $90^\circ$ scattering. Interestingly, for even smaller amplitudes we get two-bounce scattering. This is because the vortices pass the $d=0$ point and scatter under the right angle. Now, the second mode changes to the third one. Hence, its frequency still grows. Therefore the repulsion changes into the attraction between the vortices which can win over the kinetic motion. The vortices stop and move back colliding the second time. As they pass through $d=0$ the modes cross once again. The frequency drops with the separation and the interaction becomes repulsive. Thus the vortices separate indefinitely. For sufficiently small amplitude of the upper mode the kinetic energy is big enough to win over the initial vortex-vortex repulsive and later vortex-vortex attraction. It results in the usual $90^\circ$ one-bounce scattering. 

However, all these scenarios ignore the fact that for $|d|=d_*$, the third mode enters the continuum. As we have said this matters  even though we initially do not excite the third mode. After the transition of the collision point $d=0$, the second mode continues as the third mode. Thus, if the amplitude of the initially excited second mode is small enough the kinetic energy of the vortices can overcome the attractive force appearing after the first bounce and the excited 2-vortex may reach the point where the excited mode hits the mass threshold. 

In (1+1) dimensions transition of a mode, hosted on a soliton, to the continuum spectrum gives rise to the previously mentioned spectral wall phenomenon. 
\section{\label{sec:dynamics} Dynamics with the second mode excited}
\begin{figure}
	\includegraphics[height=5cm,width=8cm]{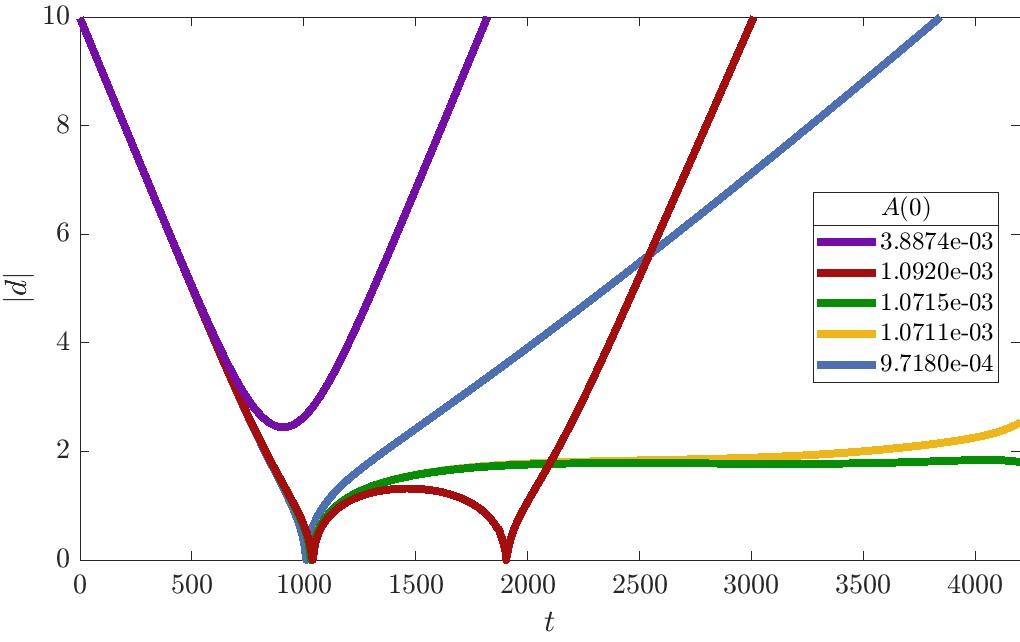}
 \caption{Dynamics of excited 2-vortex with $v_{in}=0.01$. Time evolution of $|d|$ (half vortex-vortex separation).}  
 \label{fig:dynamics}
 \vspace*{0.3cm}
		\includegraphics[height=5cm,width=8cm]{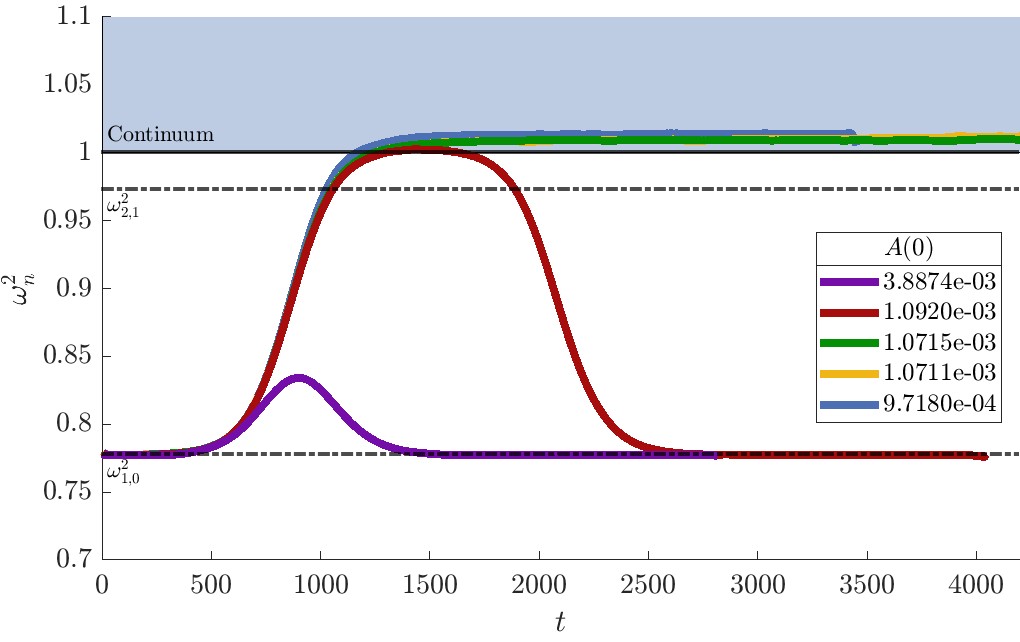}
  \caption{Time evolution of the frequency of the excited higher mode.}  
  \label{fig:modes}
  \vspace*{0.3cm}
  \includegraphics[height=5cm,width=8cm]{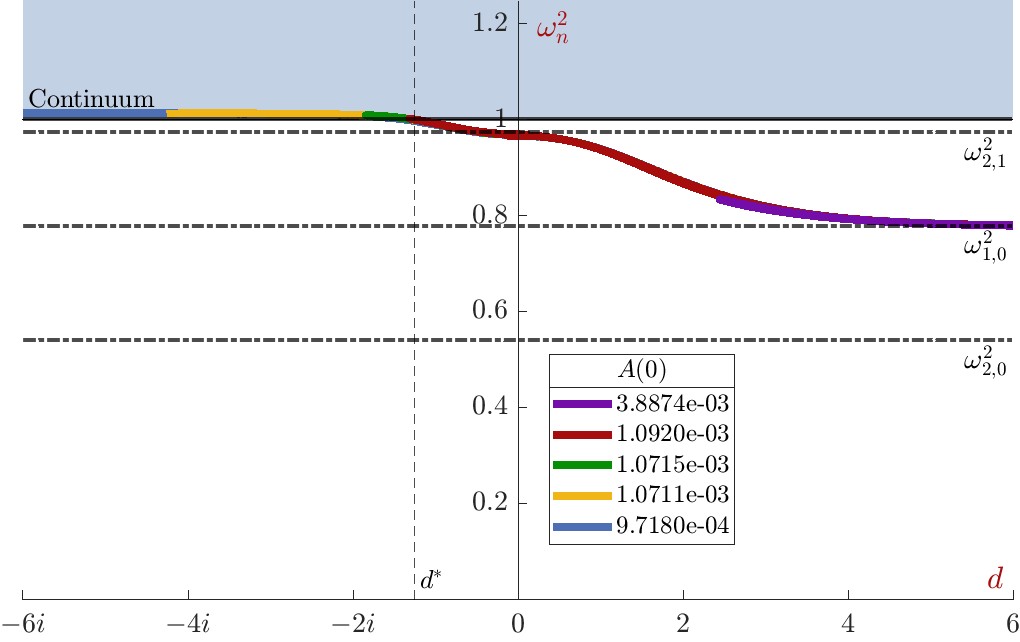}
	\caption{Frequency of the excited higher mode as a function of the vortex position parameter $d \in \mathbb{R}_+ \cup i \mathbb{R}_-$.}  
 \label{fig:mode_spe}
\end{figure}

In our numerical experiments we scatter two well separated single vortices.
Initially they are located at $x=\pm 10$ ($d(0)=10$) and boosted towards each other along $x$-axis with initial velocity $v_{in}$. The initial amplitude of the normalized mode is $A(0)$. Numerical techniques are outlined in \cite{Rees}.

In Fig. \ref{fig:dynamics} we present time evolution of $|d|$ (half of the distance between the vortices) for different values of the initial amplitude. We find the separation by tracking the zero's of the Higgs field $\phi$ by finding the minima of $|\phi|^2$ and fitting a degree $2$ polynomial to interpolate the zeros. The initial velocity is $v_{in}=0.01$. The full field theory dynamics confirms the previous considerations. For a large initial amplitude, e.g., $A(0) \approx 3.8874\mathrm{e}{-03}$ the vortices never meet. They do meet if $A(0)$ decreases. For $A(0)=1.092\mathrm{e}{-03}$ we find a two bounce solution resulting in $180^\circ$-scattering of the incoming vortices. 

As the initial amplitude further decreases the vortices can more and more separate after the first bounce. As explained, due to the mode crossing, the second mode becomes the third one. Its frequency tends to the mass threshold as the distance between the vortices along the $y$-axis increases. The spatial point at which the mode enters the continuum spectrum plays a role of a barrier in the solitonic dynamics and such a barrier is called {\it spectral wall}. 
We clearly see such a spectral wall in Fig. \ref{fig:dynamics}. The trajectory of $|d|$ gets flattened as $A(0)\approx 1.071\mathrm{e}{-03}$. This means that for a very long time the vortices almost stop at with their centers located at $y=\pm |d|_{sw}$, where $|d|_{sw}\approx 1.7$, forming a quasi-stationary state. It should be stressed that they remain at the same positions for remarkable long time. Namely, $t\approx 3000$ which can be compared with the time scale provided by the oscillation period of the excited mode. If we further reduce the initial amplitude of the mode than the kinetic energy allows to pass the spectral wall - exactly as in the case of kinks in (1+1) dimensions. Thus, for smaller $A(0)$, we enter an adiabatic regime where the standard geodesic motion of the vortices is only weakly perturbed by the excited mode. Here we encounter the usual one-bounce with $90^\circ$ scattering, see trajectory with $A(0)=9.718\mathrm{e}{-04}$. 

In Fig. \ref{fig:modes} we also plot the time evolution of the frequency of the second mode. Its behaviour very well agrees with the time evolution of the vortex-vortex separation. To numerically track the frequency, we calculate the static potential energy using Simpsons 3/8 rule at each time step $\Delta t$. We hence calculate the angular frequency, $\omega = \frac{\pi}{T_i}$, where $T_i$ denote the periods of the oscillations in the static energy. As the vortices approach each other the frequency of the mode grows. It continues growing after the vortices make the first bounce where they are on top of each other. Then they separate under $90^\circ$ angle and go along $y$-axis with the frequency still increasing. At some point the frequency equals the mass threshold, $\omega=1$. This point $d_{\omega=1}\approx  \pm 1.25$, within our numerical accuracy, coincides with the location of the stationary state, seen in Fig. \ref{fig:mode_spe}. 

In Fig. \ref{fig:mode_spe} we show how the frequency of the exited mode depends on the position $d$ of the vortex. It is worth to underline that the relation between the measured frequency and the vortex-vortex distance perfectly agrees with the results obtained by solving the linear perturbation problem, Fig. \ref{fig:espectra}.

We also observed that for higher initial velocities there is a tendency for the vortices to form the stationary solution for slightly more negative imaginary $d$ (bigger separation). This is not a surprising effect. Higher $v_{in}$ requires bigger amplitude of the mode to form the stationary solution. This means that corrections from higher order perturbation theory can matter. For example some couplings between the modes may be important. This obviously can affect the position of the spectral wall. Similar effects were observed in (1+1) dimensions \cite{SW-2, SW-3}. 
\section{\label{sec:conclusions} Conclusions}

In the current work we have shown that spectral wall, that is a barrier in solitonic motion due to transition of a bound mode to the continuum spectrum, exists also in higher dimensional BPS theories. In particular, we found it in the head-on collisions of the 2-vortex solution of the Abelian Higgs model at the critical coupling. It should be stressed that the appearance of the spectral wall in 2-vortex dynamics is a very nontrivial fact which combines the spectral structure (structure of the Hessian operator) with geometry of the moduli space. Namely, it occurs {\it after} the first collision. Therefore, it relies not only on the existence of a mode which disappears into the continuum spectrum but also on the mode crossing, which on the other hand, is strongly related to a double-covering structure of the moduli space. 

Our finding is important for several reasons.

First of all, it underlines the role of internal modes in dynamics of higher dimensional solitons. The Abelian Higgs vortices, in generality, and their collisions in particularity, have been studied extensively for decades. It is a prototypical solitonic theory in higher dimensions with obvious relation to the Standard Model and with many phenomenological applications, see e.g., superconductors or cosmic strings. However, basically all mathematical as well as numerical results concerned the limit where no internal modes were excited. We clearly show that excitation of modes may very strongly affects the dynamics, changing completely its qualitative picture. This has been known in (1+1) dimensions but now it is obvious that it also applies to physically more relevant higher dimensional models. 

Secondly, the existence of spectral walls in higher dimensions may be important in some physical applications of vortices (and other solitons). For example, if extended to the third dimensions the vortices give rise to cosmic strings. There are no reasons to expect that they could be produced in the unexcited state. Of course, ensemble of excited strings evolves differently from unexcited one \cite{V}. The number of collisions, crossings and net decay will be different - especially as the shape mode of the single vortex decays rather slowly \cite{AJJ}. This may be of some importance for the gravitational wave background generated by the cosmic strings \cite{BP}. 

Finally, looking from a wider perspective, it shows that there is a striking similarity in dynamics of the simplest topological solitons, that is one-dimensional kinks, and their higher dimensional counterpart. All the kinks do the higher dimensional solitons do as well. Thus, (1+1) dimensional solitons may be truly treated as a good laboratory for studying solitons in higher dimensions. 

There are various directions in which our work can be continued. 
One may try to identify spectral walls in other higher dimensional BPS set-ups. It means e.g., scattering of excited BPS vortices in higher topological sectors, $n>2$. One can also consider other BPS theories with planar solitons, see e.g., \cite{JW, BS, MN, TW, AFG, AK, CKM, AQW, SW, BCM}. The only condition is that the BPS vortices support bound modes, which, as flowing over the moduli space, hit the mass threshold. 

We expect that spectral walls will exist in the case of BPS solitons with a different topology. The most exciting candidates are of course the BPS monopoles. It is known that the spherical symmetric BPS monopoles support infinitely many normal \cite{Bais} (and quasinormal modes \cite{FV}), see also \cite{RS}. This is due to the fact the the gauge field remains massive also in the self-dual limit. Again, some of the normal modes can hit the mass threshold of the gauge field as we change the distance between the constituent monopoles. This will trigger the spectral wall phenomenon. 

One can also expect that spectral walls survive even if we departure a bit from the BPS regime. Thus, it would be very interesting to repeat our investigation for the non-critical coupling limit. It is known from (1+1) dimensions that sometimes appearance of a very small inter-soliton force may in fact improve the visibility of the spectral wall \cite{SW-3}. Therefore spectral walls may exist in experimentally more realistic regime both in type I and type II superconductors. This especially may concern vortices in Dirac materials, which experimentally realized in graphene, which are known to have relativistic dispersion relation, see \cite{WBB} for a review.

\section*{Acknowledgements}
This research was supported by the Spanish MCIN with funding from European
Union NextGenerationEU (PRTRC17.I1) and Consejeria de Educacion from
JCyL through the QCAYLE project, as well as the MCIN project
PID2020-113406GB-I0. Part of this research has made use of the
high-performance computing resources of the Castilla y Le\'on
Supercomputing Center (SCAYLE), financed by the European Regional
Development Fund (ERDF). AW was supported by the Polish National Science Center, 
grant NCN 2019/35/B/ST2/00059. 
Part of this research has made use of the high performance computing systems provided by the university of kent. 
MR acknowledges the UK Engineering and Physical Sciences Research Council (EPSRC) for a PhD studentship.

\subsection*{A. Movies}

Here we show movies with time dynamics of the 2-vortices with the previously defined initial conditions. Black dots are the zeros of the Higgs field. The color contours are the lines of a constant energy density. 
\begin{itemize}
\item \href{https://arxiv.org/src/2406.05725v1/anc/A09718.avi}{\texttt{A09718.avi}}
Initial amplitude of the second mode $A(0) = 9.718\mathrm{e}{-04}$. We see that the amplitude is too small, and the vortices scatter as normal, with a slight change in velocity after scattering.
\item \href{https://arxiv.org/src/2406.05725v1/anc/A10711.avi}{\texttt{A10711.avi}}
Initial amplitude of the second mode $A(0) = 1.0711 \mathrm{e}{-03}$. We see after the first collision that the vortices stop at a certain separation for a very long time, forking a very long living stationary state, but near the end of the simulation the vortices appear to begin to escape.
\item \href{https://arxiv.org/src/2406.05725v1/anc/A10715.avi}{\texttt{A10715.avi}} Initial amplitude of the second mode $A(0) = 1.0715 \mathrm{e}{-03}$. We clearly see that the vortices, after the first collision, stop at a certain separation for a very long time forming a very long living stationary state. 
\item \href{https://arxiv.org/src/2406.05725v1/anc/A10920.avi}{\texttt{A10920.avi}} Initial amplitude of the second mode $A(0) = 1.0920 \mathrm{e}{-03}$. The amplitude is too high and the vortices are reflected slightly before the spectral wall.
\item \href{https://arxiv.org/src/2406.05725v1/anc/A38874.avi}{\texttt{A38874.avi}}
Initial amplitude of the second mode $A(0) = 3.8874 \mathrm{e}{-03}$. The amplitude is too high so that the long-range forces are dominant, and the vortices repel before scattering.
\end{itemize}


\begin{thebibliography}{99}

\bibitem{B} E.~B. Bogomolny, {\it The stability of classical
    solutions}, Sov. J. Nucl. Phys. {\bf 24}, 449 (1976).
  
\bibitem{PS} M.~K. Prasad and C.~M. Sommerfield, {\it Exact
    classical solution for the 't Hooft monopole and the Julia--Zee
    dyon}, Phys. Rev. Lett. {\bf 35}, 760 (1975).

    \bibitem{Ma2} N.~S. Manton,
{\it The force between 't Hooft-Polyakov monopoles},
Nucl. Phys. \textbf{B126}, 525 (1977).

\bibitem{R} P.~J. Ruback, {\it Vortex string motion in the
    Abelian Higgs model}, Nucl. Phys. {\bf B296}, 669 (1988).
    
\bibitem{Samols} T.~M. Samols, {\it Vortex scattering},
Commun. Math. Phys. {\bf 145}, 149 (1992).
    
\bibitem{Ma1} N.~S. Manton, 
{\it A remark on the scattering of BPS monopoles},
Phys. Lett. \textbf{B110}, 54 (1982).

    
\bibitem{Rees} S. Krusch, M. Rees, T. Winyard, {\it Scattering of Vortices with Excited Normal Modes}, arXiv:2406.04164.  

\bibitem{AMMW} A. Alonso Izquierdo, N.S. Manton, 
  J. Mateos Guilarte, A. Wereszczynski, {\it Collective Coordinate Models for 2-Vortex Shape Mode Dynamics}, arXiv:2405.20249.   
    
\bibitem{sug}  T. Sugiyama, {\it Kink-antikink collisions in the two-dimensional $\phi^4$ model}, Prog. Theor. Phys. {\bf 61} (1979) 1550.

\bibitem{CSW}  D. K. Campbell, J. F. Schonfeld, and C. A. Wingate,
{\it Resonance structure in kink-antikink interactions in $\phi^4$
theory}, Physica {\bf D9} (1983) 1.

\bibitem{MORW} N. S. Manton, K. Oles, T. Romanczukiewicz, and A.
Wereszczynski, {\it Collective coordinate model of kink-antikink collisions in $\phi^4$ theory}, Phys. Rev. Lett. {\bf 127}, 071601 (2021).

 \bibitem{AORW} C. Adam, K. Oles, T. Romanczukiewicz, and A. Wereszczynski, {\it Spectral Walls in Soliton Collisions}, Phys.
Rev. Lett. {\bf 122}, 241601 (2019).

\bibitem{SW-1} C. Adam, K. Oles, T. Romanczukiewicz, A. Wereszczynski, W. Zakrzewski, {\it Spectral walls in multifield kink dynamics}, J. High Energy Phys. {\bf 2108} (2021) 147.

\bibitem{SW-2} C. Adam, K. Oles, T. Romanczukiewicz, A. Wereszczynski, {\it Spectral walls in antikink-kink scattering in the $\phi^6$ model}, Phys. Rev. {\bf D106} (2022) 105027.

 \bibitem{T} C.~H. Taubes, {\it Arbitrary $N$-vortex solutions to
the first order Ginzburg--Landau equations},
Commun. Math. Phys. {\bf 72}, 277 (1980).


\bibitem{AGMM} A. Alonso Izquierdo, W. Garcia Fuertes, N.S. Manton and
  J. Mateos Guilarte, {\it Spectral flow of vortex shape modes over
    the BPS 2-vortex moduli space},
    J. High Energy Phys. {\bf 01} (2024) 020.
    
    \bibitem{H} M. Goodband and M. Hindmarsh, {\it Bound states
and instabilities of vortices}, Phys. Rev. {\bf D52}, 4621 (1995).

\bibitem{AGG} A. Alonso-Izquierdo, W. Garcia Fuertes and
  J. Mateos Guilarte, {\it A note on BPS vortex bound states},
  Phys. Lett. {\bf B753}, 29 (2016).

\bibitem{AGG1} A. Alonso-Izquierdo, W. Garcia Fuertes and
  J. Mateos Guilarte, Dissecting zero modes and bound states on BPS vortices in Ginzburg-Landau superconductors, J. High Energy Phys. {\bf 05} (2016) 074.

\bibitem{SW-3} C. Adam, K. Oles, T. Romanczukiewicz, and A. Wereszczynski, {\it Kink-antikink collisions in a weakly interacting $\phi^4$ model}, Phys. Rev. {\bf E102}, 062214 (2020). 

\bibitem{V} D. Matsunami, L. Pogosian, A. Saurabh, and T. Vachaspati, {\it Decay of Cosmic String Loops Due to Particle Radiation} Phys. Rev. Lett. {\bf 122} 201301 (2019). 

\bibitem{AJJ} A. Alonso-Izquierdo, J.J. Blanco-Pillado, D. Miguélez-Caballero, S. Navarro-Obregón, J. Queiruga, {\it Excited Abelian-Higgs vortices: decay rate and radiation emission}, arXiv:2405.06030. 

\bibitem{BP} LISA Cosmology Working Group, Jose J. Blanco-Pillado et. al., {\it Gravitational waves from cosmic strings in LISA: reconstruction pipeline and physics interpretation}, arXiv:2405.03740.

\bibitem{JW} R. Jackiw and Erick J. Weinberg, {\it Self-dual Chern-Simons vortices}, Phys. Rev. Lett. {\bf 64}, 2234 (1990).

\bibitem{BS} B. J. Schroers, {\it Bogomol’nyi solitons in a gauged $O(3)$ sigma model}, Phys. Lett. {\bf B356}, 291 (1995).

\bibitem{MN} M. Nitta and W. Vinci, {\it Decomposing instantons in two dimensions}, J. Phys. {\bf A45}, 175401 (2012).

\bibitem{TW} D. Tong, K. Wong, {\it Vortices and Impurities}, J. High Energy Phys. {\bf 1401} (2014) 090.

\bibitem{AFG} A. Alonso-Izquierdo, W. Garcia Fuertes and J. Mateos Guilarte, {\it Two species of vortices in massive gauged non-linear sigma models}, J. High Energy Phys. {\bf 02}, 139 (2015).

\bibitem{AK} J. Ashcroft and S. Krusch, {\it Vortices and magnetic impurities},  Phys. Rev. {\bf D101} (2020) 2, 025004.

\bibitem{CKM} A. Cockburn, S. Krusch, and A. A. Muhamed, {\it Dynamics of vortices with magnetic impurities}, J. Math. Phys. {\bf 58} (2017) 6, 063509.  

\bibitem{AQW} C. Adam, J. M. Queiruga, and A. Wereszczynski, 
BPS soliton-impurity models and supersymmetry, J. High Energy Phys. {\bf 07}, 164 (2019).

\bibitem{SW} M. Speight and T. Winyard, {\it Intervortex forces in competing-order superconductors}, Phys. Rev. {\bf B103}, 014514 (2021).

\bibitem{BCM} D. Bazeia, J. G. F. Campos, and A. Mohammadi, {\it Abelian Chern-Simons vortices in the presence of magnetic impurities}, arXiv:2404.11694.

\bibitem{Bais} F. A. Bais and W. Troost, {\it Zero modes and bound states pf supersymmetric monoipole}, Nucl. Phys. {\bf B178} (1981) 125. 

\bibitem{FV} P. Forgacs and M. S. Volkov, {\it Resonant excitations of the’tHooft-Polyakov monopole}, Phys. Rev. Lett. {\bf 92} (2004) 151802.

\bibitem{RS} K. M. Russell and B. J. Schroers, {\it On resonances and bound states of the ’t Hooft-Polyakov monopole}, Phys. Rev. {\bf D83} (2011) 065004. 

\bibitem{WBB} T. O. Wehling, A. M. Black-Schaffer, and A. V. Balatsky, {\it Dirac materials}, Adv. Phys. {\bf 76}, 1 (2014). 

\end{thebibliography}
\end{document}